\documentclass[12pt]{article}

\usepackage{times}
\usepackage{geometry}
\usepackage{graphicx} 
\usepackage{amsmath}
\usepackage{subcaption}
\usepackage{amsfonts}
\usepackage{url}
\usepackage{amsthm}
\theoremstyle{definition}
\newtheorem{proposition}{Proposition}
\usepackage{xcolor}
\usepackage{tabularx}
\usepackage{array}
\usepackage{hyperref}

\begin{document}

\begin{center}
{\large\textbf{Target-Aware Network Dismantling with Limited Budgets}}\\[0.3cm]

{\normalsize
Jiaqi Song$^{1,2,3}$, Fei Jing$^{3,7}$, Wenjie Huang$^{3,4}$, Xingqin Qi$^{1,*}$, Qingpeng Zhang$^{3,5,7,8*}$}\\[0.15cm]

\begin{center}
\small

\makebox[\textwidth][c]{%
$^{1}$ School of Mathematics and Statistics, Shandong University, Weihai, China
}\\

\makebox[\textwidth][c]{%
$^{2}$ Department of Data Science, City University of Hong Kong, Hong Kong SAR, China
}\\

\makebox[\textwidth][c]{%
$^{3}$ Musketeers Foundation Institute of Data Science, The University of Hong Kong, Hong Kong SAR, China
}\\

\makebox[\textwidth][c]{%
$^{4}$ Department of Data and Systems Engineering, The University of Hong Kong, Hong Kong SAR, China
}\\

\makebox[\textwidth][c]{%
$^{5}$ HKU Shanghai Intelligent Computing Research Center, Shanghai, China
}\\

\makebox[\textwidth][c]{%
\shortstack[c]{
$^{6}$ Department of Pharmacology and Pharmacy, Li Ka Shing Faculty of Medicine,\\
The University of Hong Kong, Hong Kong SAR, China
}
}\\

\makebox[\textwidth][c]{%
$^{7}$ The University of Hong Kong -- Shenzhen Hospital, Shenzhen, China
}\\

\makebox[\textwidth][c]{%
$^{8}$ Shenzhen Loop Area Institute, Shenzhen, China
}\\[0.15cm]

$^{*}$Corresponding authors: qpzhang@hku.hk (Q.Z.), qixingqin@sdu.edu.cn (X.Q.) 

\end{center}

\end{center}

\section*{Abstract}
During epidemics, limited resources dictate critical decisions about whom to isolate or monitor. Conventional public health strategies intuitively prioritize known infected individuals and their immediate contacts. Here, we reveal that this contact-centric approach is fundamentally suboptimal under realistic resource constraints. By formulating this challenge as target-aware network dismantling, we uncover a counter-intuitive, budget-dependent intervention transition. Generating-function analysis demonstrates that when intervention budgets are scarce, isolating structurally central individuals—rather than the immediate neighbours of the infected—protects a vastly larger population. The conventional neighbour-targeting strategy only becomes optimal as available resources increase. Because this critical transition point shifts dynamically with the intervention budget, network topology, and spatial distribution of infections, static strategies inevitably fail. To overcome this, we develop Risk-Aware Isolation Learning (RAIL), a deep reinforcement learning framework that adaptively selects optimal interventions based on the evolving residual network. Across diverse synthetic and real-world networks, RAIL consistently maximizes population protection and suppresses outbreak spread. Ultimately, our findings challenge prevailing epidemiological intuitions, establishing a generalizable framework for adaptive interventions against localized risks in public health and other complex systems.


\section{Introduction}
During an epidemic, infected or exposed individuals can often be identified through testing and contact tracing~\cite{hu2021infectivity, raymenants2022empirical, ferretti2020quantifying}. However, exhaustively isolating these individuals and their contacts frequently exceeds available public health capacity~\cite{contreras2021challenges, scarselli2021discontinuous,adam2020clustering}. Limited resources must therefore be strategically allocated to maximize population protection~\cite{wallinga2010optimizing,medlock2009optimizing}. By conceptualizing populations as complex networks---where individuals and their interactions are represented as nodes and edges~\cite{salathe2010high, firth2020using}---the act of isolating an individual becomes mathematically equivalent to node removal. Consequently, this epidemiological challenge is reduced to a structural optimization problem: determining the optimal set of nodes to remove to maximally disconnect the susceptible population from known infection sources.
This framework extends well beyond epidemiology to other complex systems in which localized threats are detected before widespread diffusion. For instance, in communication and infrastructure networks, compromised servers or vulnerable facilities can often be pinpointed before systemic failures cascade~\cite{cichonski2012computer,communicationnetwork,communicationnetworks2,buldyrev2010catastrophic,infrastructurenet,infrastructurenet2}. Similarly, in online social networks, accounts propagating harmful misinformation can be flagged before the content goes viral~\cite{shu2017fake, vosoughi2018spread}. Across these diverse domains, the fundamental challenge unifies into a single, target-aware network problem: given a known set of localized risks and a strictly limited intervention budget, which nodes should be selected for removal to shield the largest possible portion of the remaining network?

Existing network interventions generally approach this challenge from three distinct, yet individually constrained, paradigms. The first is network dismantling, which employs structural heuristics~\cite{degree,freeman1977set,Sabidussi1966,Seidman1983269} or advanced learning and percolation techniques~\cite{morone2015influence,Networkdismantling,grassia2021machine,zhang2022dismantling,fan2020finding,gu2025deep} to globally fragment a system. However, these methods are fundamentally \textit{target-blind}. By optimizing for overall fragmentation, they routinely fail to explicitly disconnect the remaining network from the actual infected nodes. A second approach, rooted in influence minimization~\cite{budak2011limiting,wang2013negative,xie2023minimizing,EfficientInfluence}, actively targets known spreading sources but heavily relies on explicit diffusion models. This rigid reliance makes them fragile during the critical early stages of an outbreak, when transmission parameters remain highly uncertain~\cite{gozzi2025comparative,gibson2025improving}. Finally, a pragmatic, model-free alternative simply prioritizes the immediate neighbors of infected individuals for isolation~\cite{hellewell2020feasibility,socialcontact2}. Yet, isolating all contacts frequently exceeds practical capacities, and when budgets are strictly limited, prioritizing local neighbors inherently fails to maximize the broader protected population. Consequently, a critical gap remains: traditional frameworks force a compromise, failing to simultaneously address known localized risks, strictly limited budgets, and model-free transmission environments. The pressing challenge is how to optimally guide targeted interventions when all three of these realistic constraints coexist, necessitating a fundamentally new network formulation.

To address this, we formulate the \textit{target-aware network dismantling} problem. Instead of blindly fragmenting the entire system, this approach strategically removes a limited budget of nodes to strictly minimize the size of components that remain connected to known infection sources. By employing generating-function analysis~\cite{NetworkRobustness,newman2001random,newman2002spread}, we uncover a counter-intuitive phenomenon governing this process: in heavy-tailed networks, when the intervention budget is too small to isolate all immediate neighbors of the infected, prioritizing these neighbors is fundamentally suboptimal. Instead, removing structurally central nodes protects a significantly larger population. Crucially, we demonstrate that this relative effectiveness is strictly budget-dependent. As available resources increase, one or more critical transition points emerge where neighbor-based interventions suddenly overtake structural ones, and vice versa. Furthermore, our empirical experiments reveal that this transition point is highly volatile, shifting drastically depending on the underlying network topology and the spatial distribution of the infected nodes.

These theoretical findings expose a fundamental limitation of conventional interventions: because every node removal irrevocably alters the residual topology and reshapes the viable pathways to infection sources, any static heuristic inevitably leads to suboptimal containment. Interventions must therefore be modeled as a dynamic, sequential decision-making process driven by the continuously evolving network state. Motivated by this rigorous theoretical requirement, we develop Risk-Aware Isolation Learning (RAIL), a deep reinforcement learning framework designed explicitly for target-aware dismantling~\cite{khalil2017learning,schuetz2022combinatorial}. At each step, RAIL dynamically extracts the active residual network still connected to the infected nodes. A graph neural network encodes this shifting structural state, while a Double DQN~\cite{van2016deep} evaluates candidate interventions based on their projected long-term impact on isolating the remaining susceptible population. By selecting nodes sequentially, the learned policy generates a single, universal removal ordering, yielding flexible intervention sets that adapt seamlessly to varying budget constraints without the need for model retraining. Across diverse real and synthetic networks, RAIL consistently outperforms classical dismantling algorithms, neighbour-targeting heuristics, and adapted learning baselines. Crucially, its superior performance persists even under high uncertainty---when knowledge of the infected nodes is incomplete or noisy. Beyond structural optimization, we demonstrate that the targeted isolations prescribed by RAIL effectively halt dynamic pathogen transmission, significantly suppressing spreading in both SIR simulations and empirical outbreak trajectories on the global airport network.

\section{Preliminaries}
\subsection{Problem Formulation}

Let $G(V, E)$ be an undirected network with $N=|V|$ nodes and $M=|E|$ edges. Let $T\subseteq V$ denote the set of known infected nodes. The fraction of infected nodes is denoted by $\rho$, where $\rho=\frac{|T|}{N}$. We consider a node removal intervention, where a subset $V_R\subseteq V\backslash T$ is removed from the network. Unless otherwise specified, infected nodes are treated as known sources and are not directly removed. The removal budget is defined as $R=\frac{|V_R|}{|V\backslash T|}$. 

The classical network dismantling problem aims to identify a set of nodes $V_R\subseteq V$ to be removed such that the connectivity of the residual network $G\backslash V_R$ is minimized under a budget constraint $|V_R|\leq B$. Let $C(G\backslash V_R)$ denote the set of connected components in the residual network, and let $\delta_i=|C_i|$ denote the size of component $C_i$.
A common objective is to minimize the size of the largest connected component, which can be written as
\begin{equation}
\min_{|V_R| \leq B} \, \max_{C_i\in C(G\backslash V_R)}\delta_i
\end{equation}

Another objective of network dismantling is to minimize a connectivity measure that accounts for all connected components. A standard choice is
\begin{equation}
\min_{|V_R| \leq B} \sum_{C_i\in C(G\backslash V_R)}\frac{\delta_i(\delta_i-1)}{2}
\end{equation}

These objectives are target-blind because they just evaluate the connectivity of network without considering which components contain infected nodes.
To quantify the number of nodes that remain connected to the infected nodes after removing $V_R$, we define
\begin{equation}
    R_T(V_R)=\sum_{C_i\in C(G\backslash V_R)}\delta_i\mathbb{I}(C_i\cap T\neq\emptyset)
\end{equation}
where $\mathbb{I}(\cdot)$ is the indicator function. 

The target-aware network dismantling can then be formulated as
\begin{equation}
\min_{|V_R| \leq B} \, R_T(V_R)
\end{equation}

Unlike classical dismantling, this objective only considers nodes that remain in components containing at least one node in $T$. Removing structurally important nodes is often effective in classical network dismantling. Under the target-aware objective, however, it is natural to expect that removing neighbors of infected nodes would be more effective. It therefore remains unclear whether target-aware network dismantling benefits more from removing structurally important nodes or from removing neighbors of infected nodes.

\subsection{Effective Removal Strategy Changes with the Intervention Budget}
\begin{figure*}[t]
    \centering

    \begin{tabular}{@{}ccc@{}}

        \begin{subfigure}[b]{0.30\textwidth}
            \centering
            \includegraphics[width=\linewidth]
            {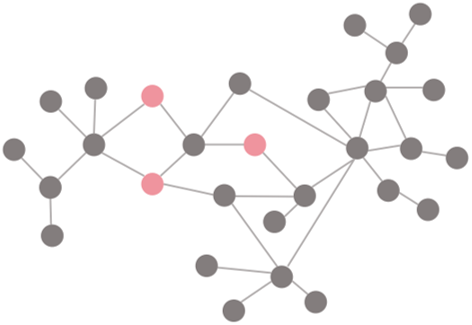}
            \caption{Original}
            \label{fig:diff_original}
        \end{subfigure}
        &
        \begin{subfigure}[b]{0.30\textwidth}
            \centering
            \includegraphics[width=\linewidth]
            {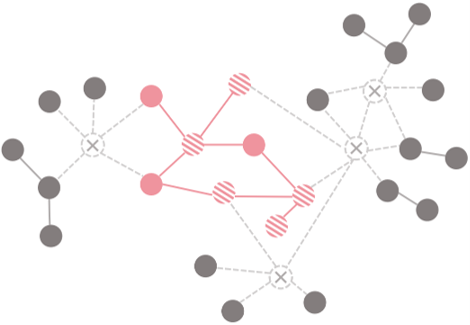}
            \caption{High-degree, $B=4$}
            \label{fig:diff_centrality}
        \end{subfigure}
        &
        \begin{subfigure}[b]{0.30\textwidth}
            \centering
            \includegraphics[width=\linewidth]
            {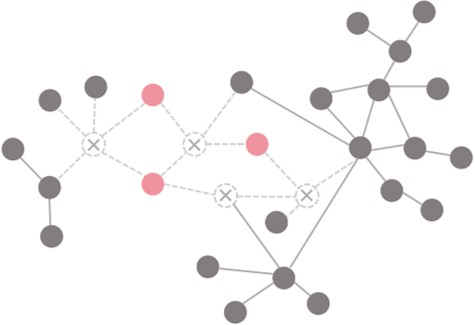}
            \caption{Neighbor-based, $B=4$}
            \label{fig:diff_target}
        \end{subfigure}

        \\[4mm]

        &
        \begin{subfigure}[b]{0.30\textwidth}
            \centering
            \includegraphics[width=\linewidth]
            {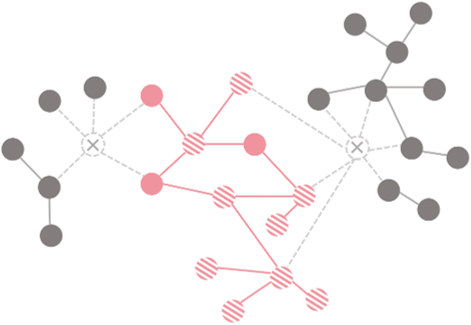}
            \caption{High-degree, $B=2$}
            \label{fig:diff_centrality_small}
        \end{subfigure}
        &
        \begin{subfigure}[b]{0.30\textwidth}
            \centering
            \includegraphics[width=\linewidth]
            {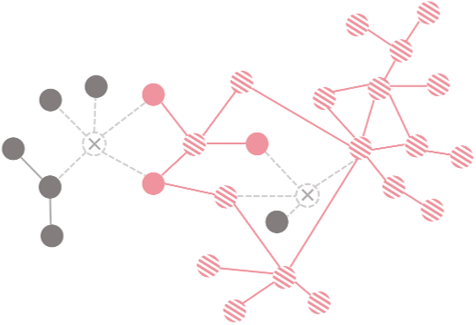}
            \caption{Neighbor-based, $B=2$}
            \label{fig:diff_target_small}
        \end{subfigure}

    \end{tabular}

    \caption{\textbf{Comparison of two common removal strategies, high-degree and neighbor-based removal, under different intervention budgets.} (a) The original network contains 29 nodes, including three infected nodes. Solid pink nodes denote infected nodes, and gray nodes denote uninfected nodes. In (b)–(e), striped pink nodes denote uninfected nodes that remain connected to at least one infected node after removal, gray nodes denote uninfected nodes that are disconnected from all infected nodes, and crossed nodes denote removed nodes. After removing four nodes, (b) high-degree removal leaves 8 nodes in components containing infected nodes, whereas (c) neighbor-based removal leaves 3. With a smaller budget of two removed nodes, (d) high-degree removal leaves 12 nodes in components containing infected nodes, whereas (e) neighbor-based removal leaves 21.}
    \label{fig:difference}
\end{figure*}

To answer this question, we first compare two intuitive removal strategies: removing high-degree nodes and removing neighbors of infected nodes. These two strategies reduce the number of nodes that remain connected to the infected nodes in different ways. Removing high-degree nodes can disconnect large parts of the network, whereas removing neighbors of infected nodes can isolate them directly. 
This distinction can be illustrated by the examples in Figure~\ref{fig:difference}.

When the budget is sufficient, removing nodes adjacent to the infected set can directly isolate the region containing infected nodes and leave fewer nodes connected to that set (Figure~\ref{fig:diff_target}). When the budget is very small, however, removing a structurally important nodes can disconnect a larger part of the network from the infected nodes than removing an immediate neighbor of infected node (Figure~\ref{fig:diff_centrality_small}). Under the target-aware network dismantling objective, these examples suggest that the relative effectiveness of high-degree removal and removing neighbors of infected nodes depends on the intervention budget. We next use generating function analysis to explain when each strategy is more effective.

\section{Theoretical Analysis and Experimental Validation}
\subsection{Theoretical Analysis}
We use generating function analysis~\cite{NetworkRobustness} to explain why removing neighbors of infected nodes is not always the most effective strategy. We compare two removal strategies: high-degree removal and neighbor-based removal, where neighbor-based removal selects nodes from the neighbors of infected nodes.

We consider a random network whose degree distribution is given by $\{p_k\}_{k\geq 0}$, where $p_k$ is the probability that a randomly selected node has degree $k$, and the average degree is $\langle k\rangle=\sum_k kp_k$. Each node is independently marked as infected with probability $\rho$. Among the uninfected nodes, a node whose degree is $k$ before intervention is retained in the residual network with probability $q_k$. The expected fraction of nodes in components containing at least one infected node can then be expressed using the degree generating function of the remaining uninfected nodes and the corresponding excess degree generating function.

Random removal is included as a special case in which $q_k$ is independent of $k$, with $q_k=1-R$ for all $k$. 
Previous studies have shown that in heterogeneous networks, targeted removal can be substantially more effective than random removal~\cite{communicationnetwork, communicationnetworks2,pastor2002immunization}. We therefore focus on high-degree removal and neighbor-based removal, which use network structure and information about infected nodes, respectively.

\begin{proposition}[Expected fraction of nodes in components containing infected nodes]\label{generate}
Under the setting described above, we define the degree generating function of the remaining uninfected nodes as

\begin{equation*}
F_0(x)=(1-\rho)\sum_k p_kq_kx^k,
\end{equation*}
and the corresponding excess degree generating function as
\begin{equation*}
F_1(x)=(1-\rho)\sum_k\frac{kp_k}{\langle k\rangle}q_kx^{k-1}.
\end{equation*}
where the excess degree is the number of remaining edges after reaching a node by following a randomly chosen edge.

Let $u$ denote the probability that the branch reached by a randomly chosen edge contains no infected node. Under the generating function approximation, $u$ satisfies
\begin{equation*}
u=(1-\rho)-F_1(1)+F_1(u).
\end{equation*}
The expected fraction of nodes that belong to connected components containing at least one infected node is then
\begin{equation*}
S=\rho+F_0(1)-F_0(u).
\end{equation*}

Equivalently, $S$ gives the normalized approximation of the target-aware objective, $S\approx \mathbb{E}[R_T]/N$.
\end{proposition}

The full derivation is provided in Supplementary Section 1. 

With the expression for $S$, we can analyze how $S$ varies as a function of the removal budget $R$. Specifically, the derivative $\frac{dS}{dR}$ measures the change in the fraction of nodes connected to at least one infected node as the removal budget increases. Since a more effective intervention produces a larger reduction in $S$, we define the marginal gain as $-\frac{dS}{dR}$. We first focus on the two limiting regimes, corresponding to very small and very large removal budgets. In particular, we examine their marginal gains as $R\to 0^+$ and $R\to 1^-$.

\begin{proposition}\label{smallbudget}
    High-degree removal has a larger marginal gain in reducing connected components containing infected nodes under very small budgets in heavy-tailed networks.
\end{proposition}

\begin{proposition}\label{largebudget}
    Neighbor-based removal has a larger marginal gain in reducing connected components containing infected nodes in very large budgets under a mean-field approximation.
\end{proposition}

The full derivations above are provided in Supplementary Section 2.

The two propositions characterize the relative advantage of the two strategies in the limits $R\rightarrow 0$ and $R\rightarrow 1$. To examine how their relative advantage changes at intermediate budgets, we define the difference in marginal gain as

\begin{equation*}
    \Delta(R)=-\frac{dS_{neighbor}}{dR}+\frac{dS_{degree}}{dR}
\end{equation*}

The sign of $\Delta(R)$ determines the relative advantage of the two strategies, with positive values indicating a larger marginal gain for neighbor-based removal and negative values indicating a larger marginal gain for high-degree removal. A transition occurs when the two strategies have the same marginal gain. Therefore, a transition point $R_c$ satisfies $\Delta(R_c)=0$. Importantly, the two limiting results do not imply that there is only one transition point. Between the small- and large-budget regimes, $\Delta(R)$ may change sign more than once. The transition points can therefore be written as $R_{c,1},R_{c,2},
\cdots$, where each $R_{c,i}$ is a solution of $\Delta(R)=0$. 

Both $u$ and $q_k$ vary with the removal budget $R$, and $u$ is determined implicitly by the generating function equation. Therefore, $R_c$ generally cannot be solved explicitly from the transition condition $\Delta(R)=0$, and deriving a closed form expression remains difficult. Determining these transition points analytically or approximately remains a direction for future work.

Since explicit expressions for these transition points are difficult to obtain, we next use experiments on finite networks to examine how the relative advantage changes across the full range of removal budgets. We consider two experimental settings: one based on the removal probabilities used in the analytical model and the other based on the actual neighbors of explicitly placed infected nodes.

\subsection{Experimental Validation}

Because the generating function analysis models intervention by assigning removal probabilities to nodes, it does not directly select the actual neighbors of the observed infected nodes. We therefore consider two intervention settings in finite networks. In the first setting, nodes are selected according to the removal probabilities used in the analytical model. In the second setting, intervention directly selects the actual neighbors of the specified infected nodes.

For the validation experiments, we use synthetic networks with $N=1000$ nodes. Unless otherwise specified, the base network is a scale-free network with power-law exponent $\gamma=2.5$, and the density of infected nodes is fixed at $\rho=5\%$. We then vary one aspect of the network or the distribution of infected nodes at a time. 
We vary both network structure and the placement of infected nodes. Degree heterogeneity is controlled by the power-law exponent $\gamma$, and community structure is controlled by the inter-community mixing parameter $\mu$, where smaller $\mu$ indicates clearer community structure. We vary the distribution of infected nodes through four factors: their density, spatial radius, number of clusters, and degree bias.
To control the degree bias, we sample node $v$ as an infected nodes with probability proportional to $k_v^\alpha$, where $k_v$ is the degree of $v$.  
Detailed procedures are provided in Supplementary Section 3.

\begin{figure}[htbp]
    \centering
    \captionsetup[subfigure]{justification=centering,singlelinecheck=false}
    
    \begin{subfigure}[t]{0.32\linewidth}
        \centering
        \includegraphics[width=\linewidth]{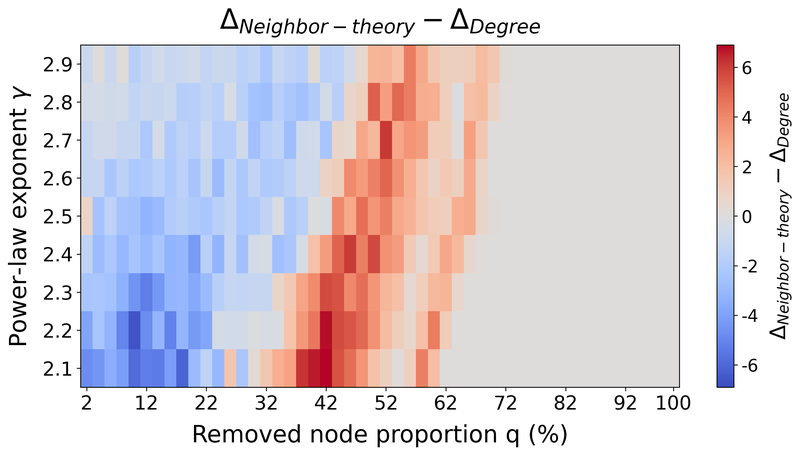}
        \caption{Degree heterogeneity}
        \label{fig:theory_powerlaw}
    \end{subfigure}
    \hfill
    \begin{subfigure}[t]{0.32\linewidth}
        \centering
        \includegraphics[width=\linewidth]{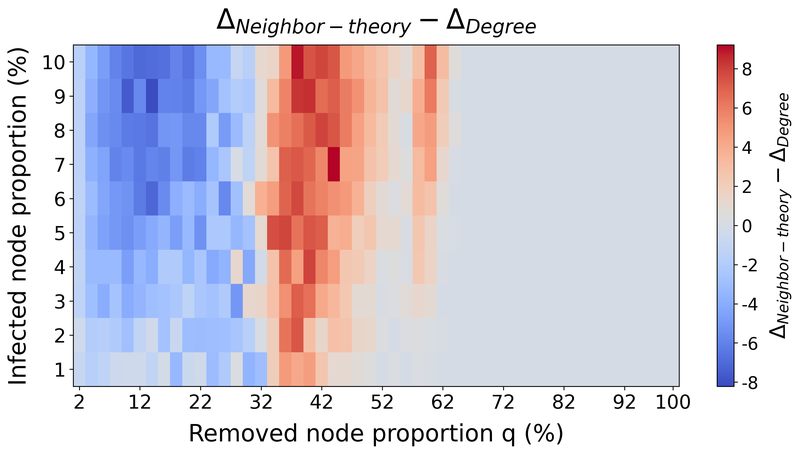}
        \caption{Density of infected nodes}
        \label{fig:theory_riskprop}
    \end{subfigure}
    \hfill
    \begin{subfigure}[t]{0.32\linewidth}
        \centering
        \includegraphics[width=\linewidth]{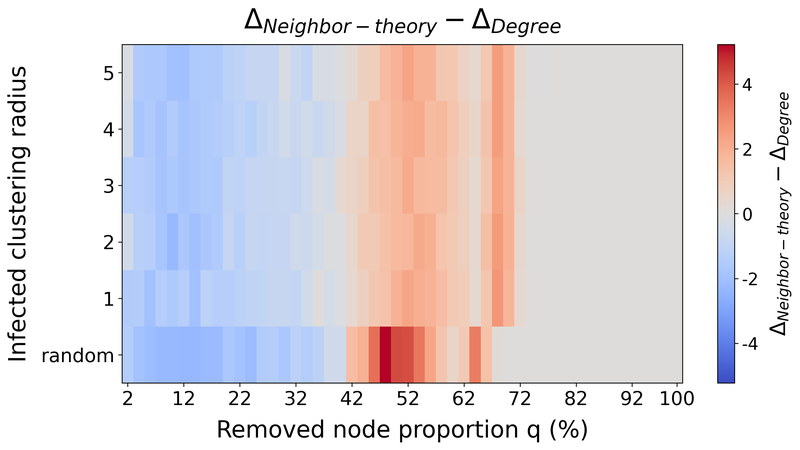}
        \caption{Spatial radius of \\ infected nodes}
        \label{fig:theory_radius}
    \end{subfigure}

    \caption{\textbf{Validation of the budget-dependent transition under analytical removal rules.} Each panel reports the difference in single-step marginal gain between high-degree and neighbor-based removal under rules matched to the generating function approximation. Red regions indicate budgets and parameter settings where neighbor-based removal gives a larger marginal reduction in the components connected to infected nodes, whereas blue regions indicate a larger marginal reduction by high-degree removal. The panels vary degree heterogeneity of the underlying network, the density and spatial radius of infected nodes.}
    \label{fig:theory}
\end{figure}

Figure~\ref{fig:theory} shows the budget-dependent transition under removal rules matched to the generating function approximation. 
The results show that high-degree removal has a larger marginal gain in small budgets in most settings, especially when the degree distribution is more heterogeneous. As the intervention budget increases, the relative advantage shifts toward neighbor-based removal. These observations match the theoretical predictions.The transition point is not fixed. It changes with degree heterogeneity of the underlying network, the density, and spatial radius of infected set.

The experiments based on the analytical rules show that neither high-degree removal nor neighbor-based removal is uniformly dominant. Their relative advantage changes with the removal budget and with the graph and settings of infected nodes. We next test whether this transition persists when directly removing neighbors of infected nodes in finite networks.

\begin{figure}[htbp]
    \centering
    \captionsetup[subfigure]{justification=centering,singlelinecheck=false}
    
    \begin{subfigure}[t]{0.32\linewidth}
        \centering
        \includegraphics[width=\linewidth]{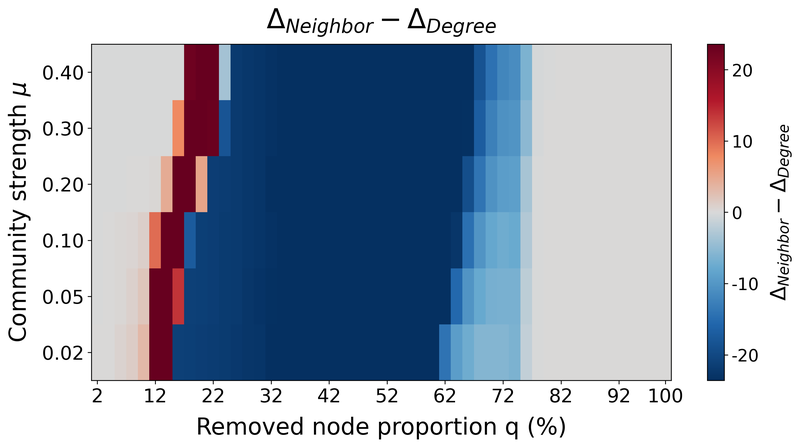}
        \caption{Community structure}
        \label{fig:real_communitynetwork}
    \end{subfigure}
    \hfill
    \begin{subfigure}[t]{0.32\linewidth}
        \centering
        \includegraphics[width=\linewidth]{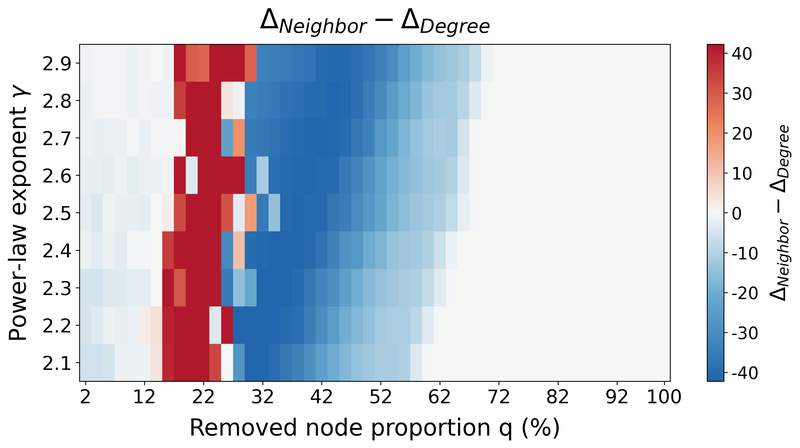}
        \caption{Degree heterogeneity}
        \label{fig:real_powerlaw}
    \end{subfigure}
    \hfill
    \begin{subfigure}[t]{0.32\linewidth}
        \centering
        \includegraphics[width=\linewidth]{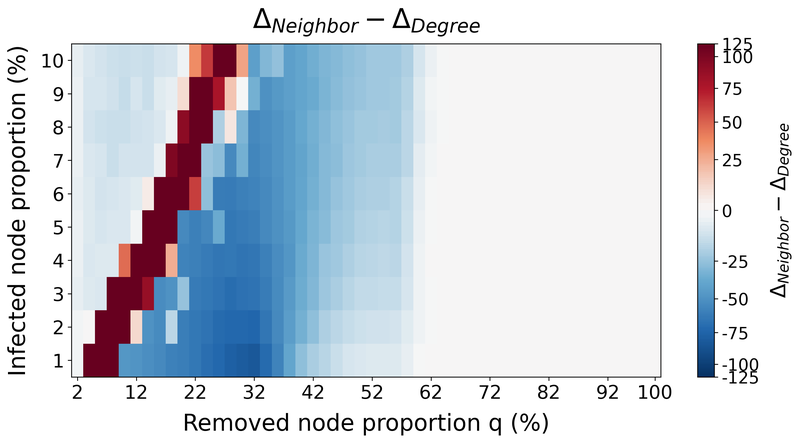}
        \caption{Density of infected nodes}
        \label{fig:real_riskprop}
    \end{subfigure}

    \vspace{1em} 
    
    \begin{subfigure}[t]{0.32\linewidth}
        \centering
        \includegraphics[width=\linewidth]{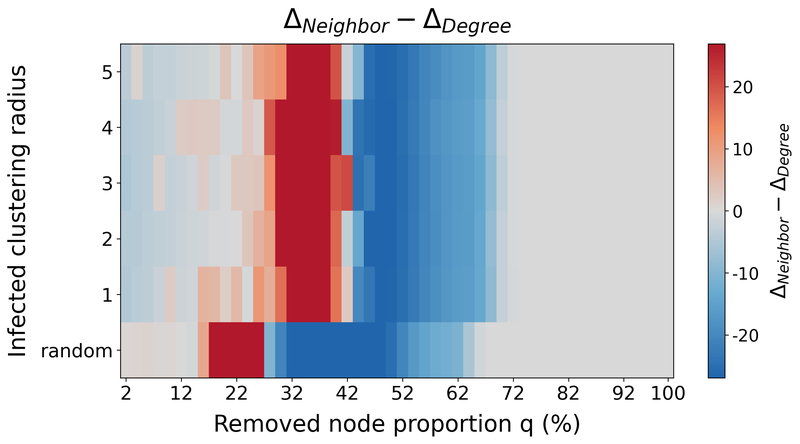} 
        \caption{Spatial radius of \\ infected nodes} 
        \label{fig:real_radius}
    \end{subfigure}
    \hfill
    \begin{subfigure}[t]{0.32\linewidth}
        \centering
        \includegraphics[width=\linewidth]{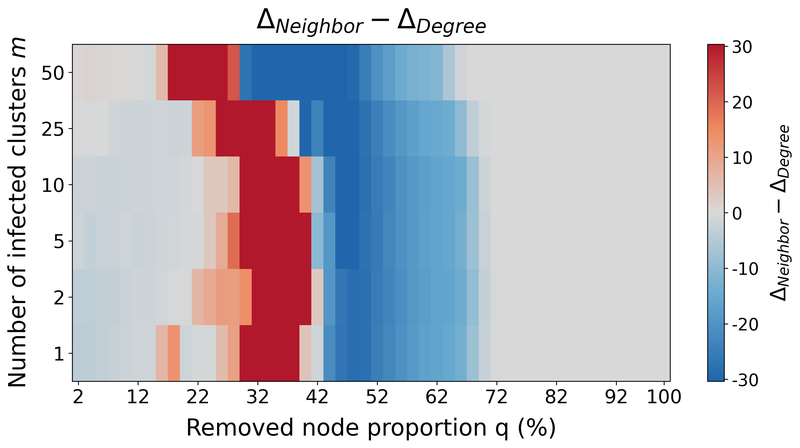} 
        \caption{Number of infected clusters} 
        \label{fig:real_clustersnumber}
    \end{subfigure}
    \hfill
    \begin{subfigure}[t]{0.32\linewidth}
        \centering
        \includegraphics[width=\linewidth]{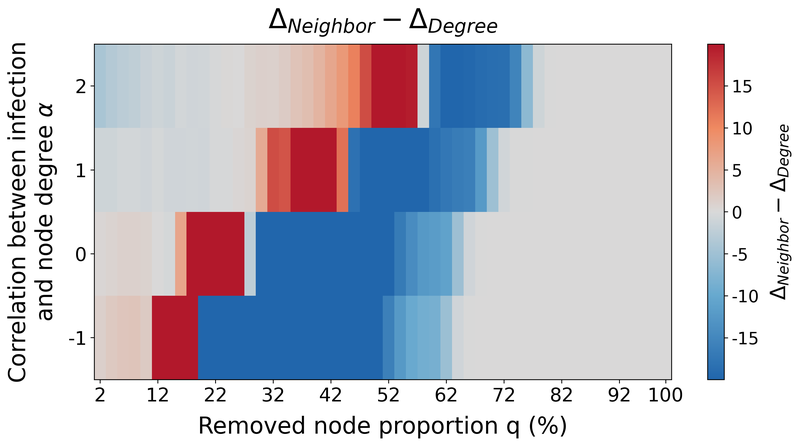} 
        \caption{Degree bias of infected nodes} 
        \label{fig:real_riskdegreebias}
    \end{subfigure}

    \caption{\textbf{Validation of the budget-dependent transition with observed neighbors of infected nodes.} The neighbor-based strategy removes actual neighbors of infected nodes in finite networks. Colors report the difference in single-step marginal gain between high-degree removal and neighbor-based removal. The panels show that the more effective rule changes with removal budget, network structure and the placement of infected nodes.}
    \label{fig:real}
\end{figure}

Figure~\ref{fig:real} shows that the budget-dependent transition persists when the neighbor-based strategy removes the actual neighbors of infected nodes. High-degree removal is often stronger at the beginning of the removal process, whereas neighbor-based removal becomes more competitive when the budget increases. In intermediate regimes, the dominant rule can alternate.

The transition is shaped by the underlying network structure. As community structure becomes stronger, neighbor-based removal becomes more effective at lower budgets. In contrast, when community structure is weaker, high-degree removal remains more effective over a wider range of budgets. Degree heterogeneity mainly amplifies the early advantage of high-degree removal. The distribution of infected nodes also changes the transition. As the proportion of infected nodes increases, the transition generally occurs at larger budgets. Changing the spatial radius does not produce a consistent shift in the transition. As the number of clusters increases, the transition tends to occur at smaller budgets. When infected nodes are more likely to have high degree, high-degree removal remains more effective over a larger range of budgets.

Ultimately, these analytical and empirical insights converge on a fundamental principle: the optimal intervention strategy is not static, but is dynamically governed by a complex interplay among the available budget, network topology, and the spatial distribution of infection sources. Consequently, relying on any single heuristic---or even a hand-crafted hybrid of structural and neighbor-based rules---inevitably yields suboptimal containment, because the optimal decision boundary constantly shifts as the network is sequentially dismantled. This inherent structural volatility mathematically necessitates an adaptive, state-dependent approach. Driven by this theoretical imperative, we introduce our deep reinforcement learning framework. Rather than committing to rigid, pre-defined rules, the proposed method continuously extracts the active residual network connected to the infected nodes, learning to map these dynamically evolving structural states to optimal intervention decisions.

\begin{figure*}[t]
    \centering
    \includegraphics[width=0.85\textwidth]{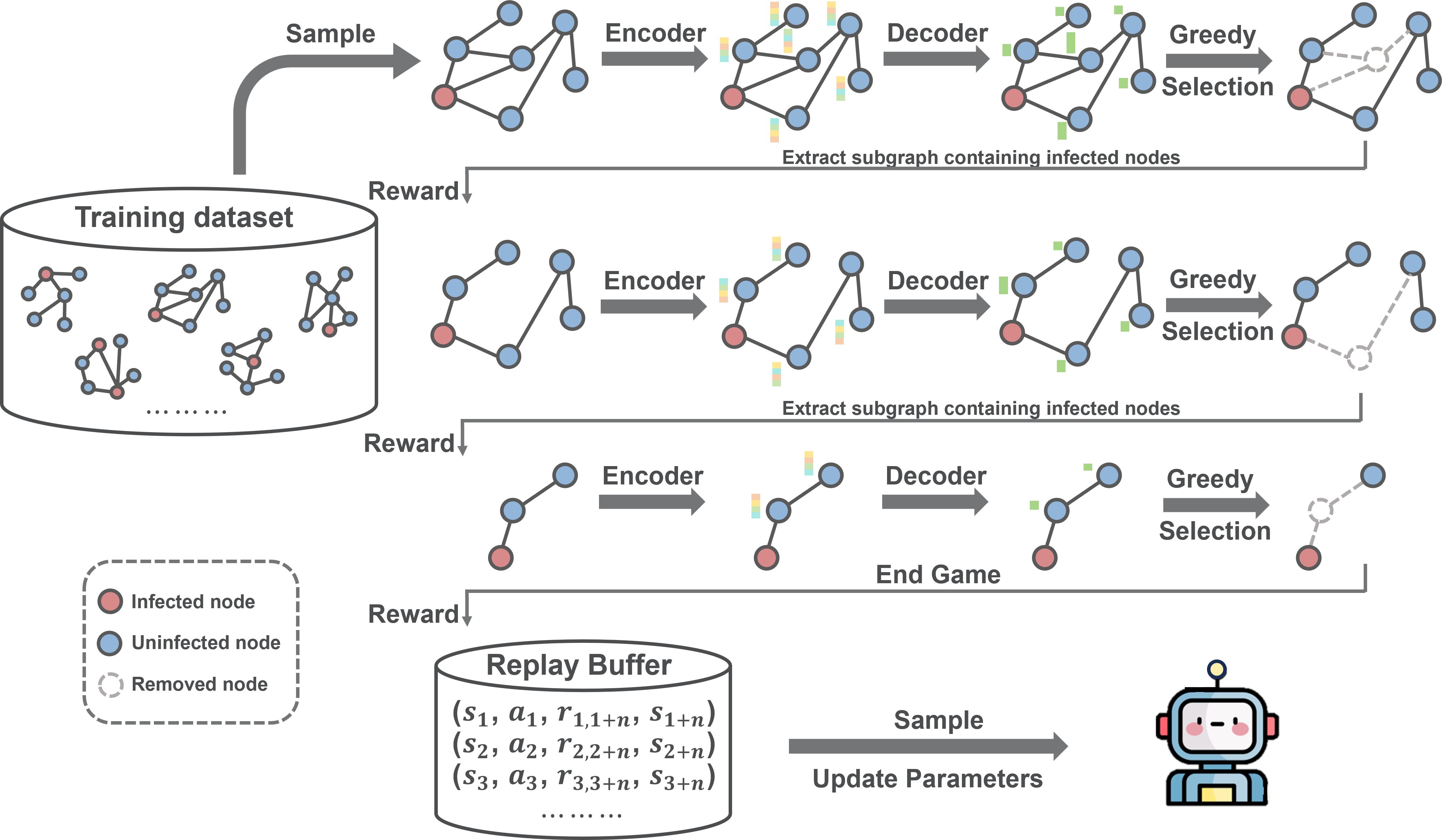}
    \caption{\textbf{Overview of RAIL.} At each removal step, RAIL extracts the residual network connected to infected nodes and encodes the current network state. The decoder evaluates candidate nodes and selects the next node for removal. The resulting transitions and rewards are stored in the replay buffer and sampled to update the model parameters during offline training.}
    \label{fig:rail_framework}
\end{figure*}

\section{Learning State-Dependent Interventions with RAIL}

To operationalize the state-dependent principles established in our theoretical analysis, we develop the Risk-Aware Isolation Learning (RAIL) framework. By synergizing a Graph Neural Network (GNN) with a Double DQN decoder, RAIL continuously tracks the structural shifts of the network following each node removal, directly translating the demand for dynamic adaptability into a robust algorithmic pipeline. To address the reality that intervention budgets fluctuate unpredictably during real-world crises, RAIL fundamentally avoids the computational bottleneck of re-optimizing for distinct budget thresholds. Instead, it generates a single, universal ordering of removable uninfected nodes. Optimal intervention sets for any specific budget constraint are elegantly derived by taking prefixes of this learned trajectory, enabling continuous policy evaluation across the entire dismantling process. While RAIL is inspired by the sequential decision-making architecture of FINDER~\cite{fan2020finding}, it fundamentally reconceptualizes the state representation, node features, candidate action space, and reward formulation to meet the unique, localized demands of target-aware dismantling. Detailed algorithmic procedures are provided in Supplementary Section 4.

\subsection{Dynamic State Encoding via Graph Neural Networks}
RAIL uses a graph neural encoder to represent both candidate nodes and the current residual network connected to infected nodes. At removal step $t$, the current residual network is first obtained by deleting all previously selected nodes. RAIL then extracts the union of components connected to at least one infected node. The initial feature vector of each node in the residual network includes its risk status. A graph-level feature is also initialized to include the current proportion of infected nodes in the subgraph. Similar to GraphSAGE~\cite{graphsage}, RAIL applies a message-passing encoder with sum aggregation to update the embeddings according to the structure of the current residual network. After several propagation layers, each node embedding captures the structural role of the node within the residual network, while the graph embedding summarizes the overall state of the subgraph. Because the residual network is recomputed after every removal, the encoder dynamically tracks the part of the network that remains connected to the infected nodes and is relevant to the objective of target-aware network dismantling.

\subsection{Action Evaluation on Evolving Topologies}
The decoder evaluates which node should be removed next. At each step, candidate actions are restricted to nodes outside the set of infected nodes that remain in the current residual network. For each candidate, RAIL combines its node embedding with the embedding of the graph to represent the current state and the candidate action. This representation is passed through a multilayer perceptron to produce a Q-value, which estimates the future benefit of removing that node from the current network. During training, actions are selected using an $\epsilon$-greedy strategy to balance exploration and exploitation. During evaluation, RAIL selects the candidate with the largest Q-value.

\subsection{Sequential Policy Optimization and Training}
RAIL is trained offline on small synthetic graphs with 30-50 nodes. We start with 1000 training graphs and regenerate the training set every 5000 iterations. The encoder and decoder are trained end to end using n-step Double DQN with experience replay. After each episode, transitions collected from the removal process are stored in a replay buffer and sampled to update the model. A structural regularization term is also added to stabilize node embeddings as the residual network connected to the infected nodes changes after each removal.

\section{Experimental Evaluation}

To comprehensively validate RAIL, we conduct three progressive tiers of evaluation. First, we assess its direct efficacy in minimizing the cumulative size of connected components that harbor infected nodes. Second, we examine whether these structural isolations effectively suppress epidemic outbreaks through extensive SIR simulations. Third, we test RAIL's real-world deployability against empirical spreading trajectories across the global airport network. Across both the component reduction and SIR evaluations, we rigorously benchmark all methods under conditions of both complete and uncertain information regarding the initial infected nodes.

\subsection{Experimental Setting}
We evaluate RAIL on real and synthetic networks from criminal, social, information, communication, transportation and biological systems. The benchmark real networks include Crime~\cite{crime}, Digg~\cite{crime}, Enron~\cite{enron}, Epinions~\cite{crime}, Facebook~\cite{crime}, Gnutella31~\cite{Gnutella31}, HI-II-14~\cite{hi}, cbn~\cite{cbn}, faa~\cite{faa}, Figeys~\cite{fey} and vidal~\cite{vital}. We also use BA, ER and WS synthetic networks to evaluate. In addition to randomly assigned infected nodes, we evaluate RAIL on two real datasets that provide predefined infected nodes: Blued and Project 90~\cite{project90}. These datasets validate the effectiveness of RAIL when infected nodes are predefined by real data. Details of the datasets and label construction are provided in Supplementary Section 5.

We compare RAIL with standard network dismantling methods, including FINDER~\cite{fan2020finding}, CI~\cite{morone2015influence}, CoreHD~\cite{corehd}, GDM~\cite{grassia2021machine}, GND~\cite{generalizednetwork}, MinSum~\cite{Networkdismantling}, and MultiDismantler~\cite{gu2025deep}, and with baselines that use information about the known infected nodes, including Target-aware FINDER and Neighbor. Neighbor removes uninfected nodes that are adjacent to infected nodes. Target-aware FINDER retains the original FINDER architecture, excludes infected nodes from candidate actions, and replaces the original reward with the reward defined by our objective. Since MultiDismantler is designed for multilayer networks and is computationally demanding on large networks, we evaluate it only on smaller networks.

All methods generate a ranked sequence of removable uninfected nodes. Unless otherwise specified, we evaluate the residual network after removing $2\%,4\%,\cdots,30\%$ of all nodes. For the larger Blued network, the ranked sequences produced by all methods are substantially shorter than $30\%$ of the network. Therefore, we instead evaluate absolute removal budgets from 1000 to 10000 nodes. Infected nodes are randomly sampled for the synthetic and real networks, whereas predefined labels indicating infected nodes are used for Blued and Project 90.

\begin{figure}[htbp]
    \centering

    \begin{subfigure}[t]{0.48\linewidth}
        \centering
        \includegraphics[width=\linewidth]{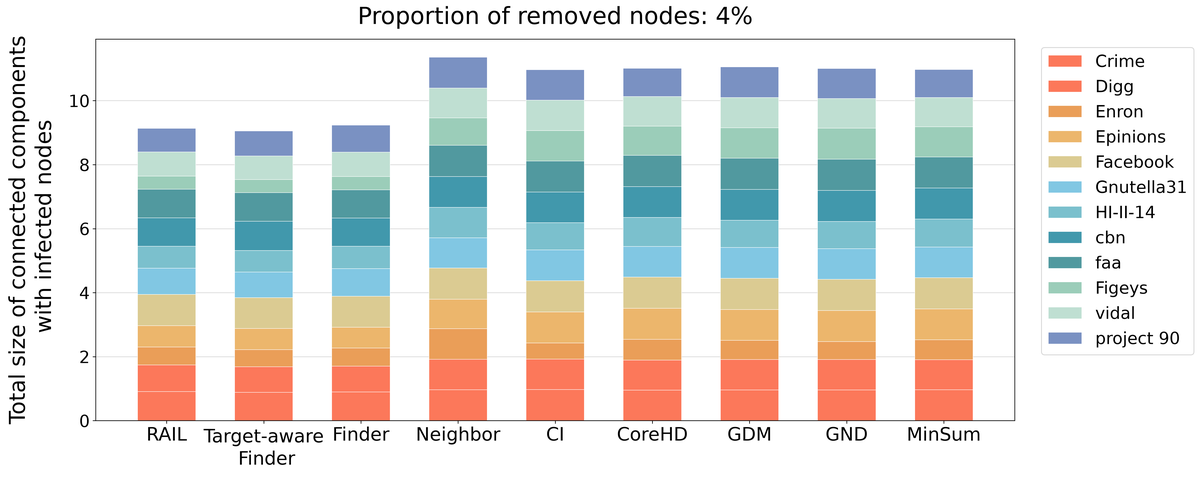}
        \caption{Removal ratio: 4\%}
        \label{fig:bar4}
    \end{subfigure}
    \hfill
    \begin{subfigure}[t]{0.48\linewidth}
        \centering
        \includegraphics[width=\linewidth]{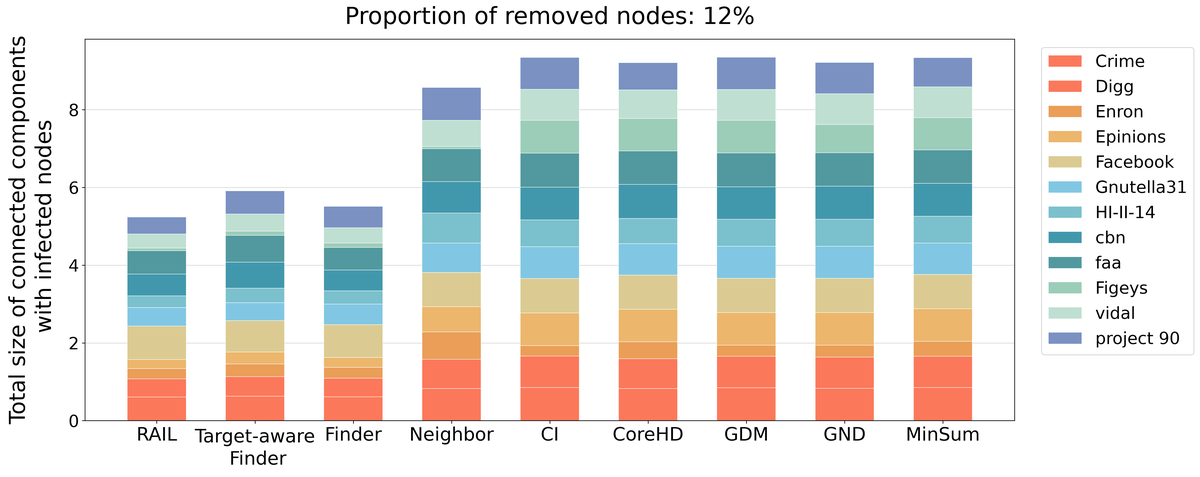}
        \caption{Removal ratio: 12\%}
        \label{fig:bar12}
    \end{subfigure}

    \vspace{0.4cm}

    \begin{subfigure}[t]{0.48\linewidth}
        \centering
        \includegraphics[width=\linewidth]{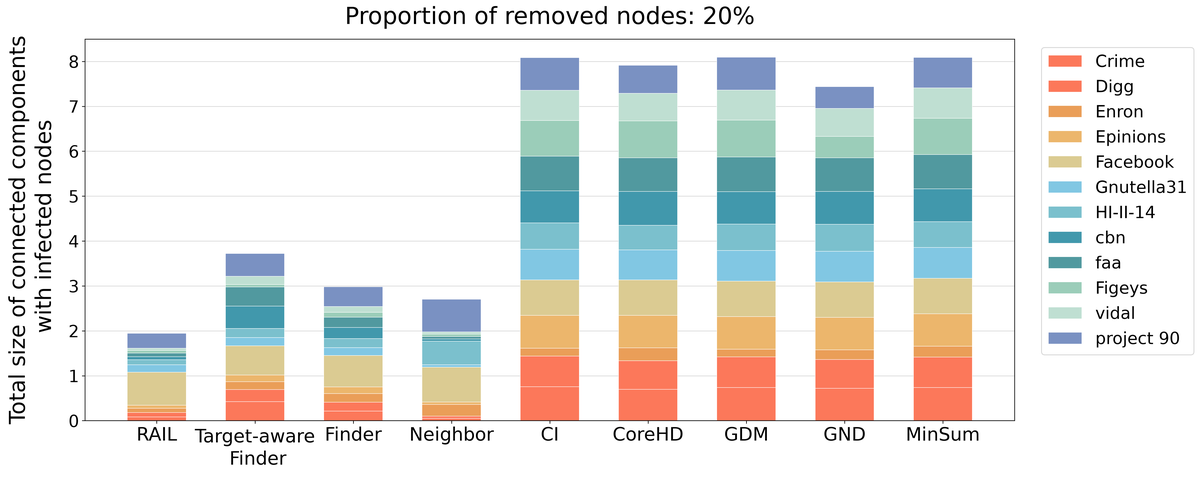}
        \caption{Removal ratio: 20\%}
        \label{fig:bar20}
    \end{subfigure}
    \hfill
    \begin{subfigure}[t]{0.48\linewidth}
        \centering
        \includegraphics[width=\linewidth]{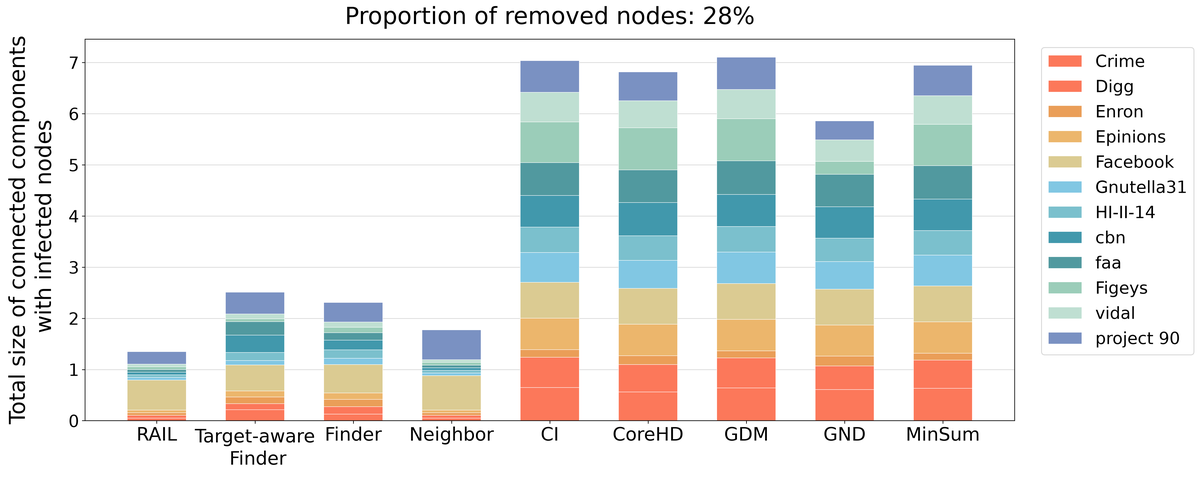}
        \caption{Removal ratio: 28\%}
        \label{fig:bar28}
    \end{subfigure}

    \caption{\textbf{Size of connected components containing at least one infected node under fully observed infected nodes.} Each panel corresponds to one removal ratio. For each method, the stacked bar aggregates the normalized $R_T(V_R)$ across all real networks. Smaller total bar height indicates better isolation of infected nodes from the remaining network.}
    \label{fig:bar}
\end{figure}

\subsection{Reducing Components Containing Infected Nodes}

We first evaluate the total size of components containing at least one infected node after node removal, which is previously denoted by $R_T(V_R)$. Smaller values indicate better intervention performance. 
The main text reports four representative removal budgets ($4\%$, $12\%$, $20\%$, and $28\%$), while complete results and experimental details are provided in Supplementary Sections 5 and 6. For each removal budget, the results are summarized using stacked bar charts. Each bar corresponds to one method, and the stacked segments represent the normalized $R_T(V_R)$ values from different networks. Therefore, a shorter total bar indicates better isolation performance.

\begin{figure}[htbp]
    \centering

    \begin{subfigure}[t]{0.48\linewidth}
        \centering
        \includegraphics[width=\linewidth]{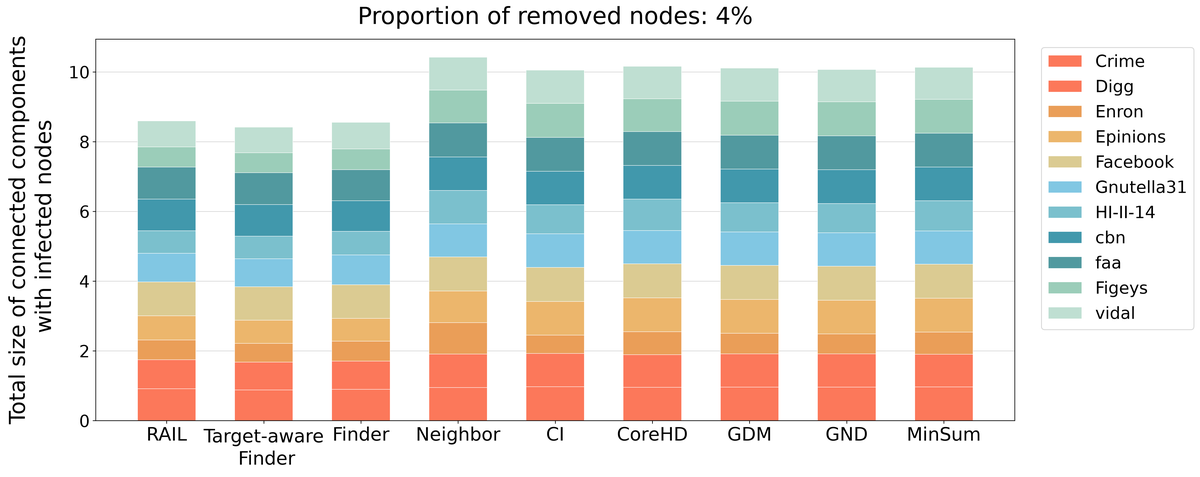}
        \caption{Removal ratio: 4\%}
        \label{fig:barfuzzy4}
    \end{subfigure}
    \hfill
    \begin{subfigure}[t]{0.48\linewidth}
        \centering
        \includegraphics[width=\linewidth]{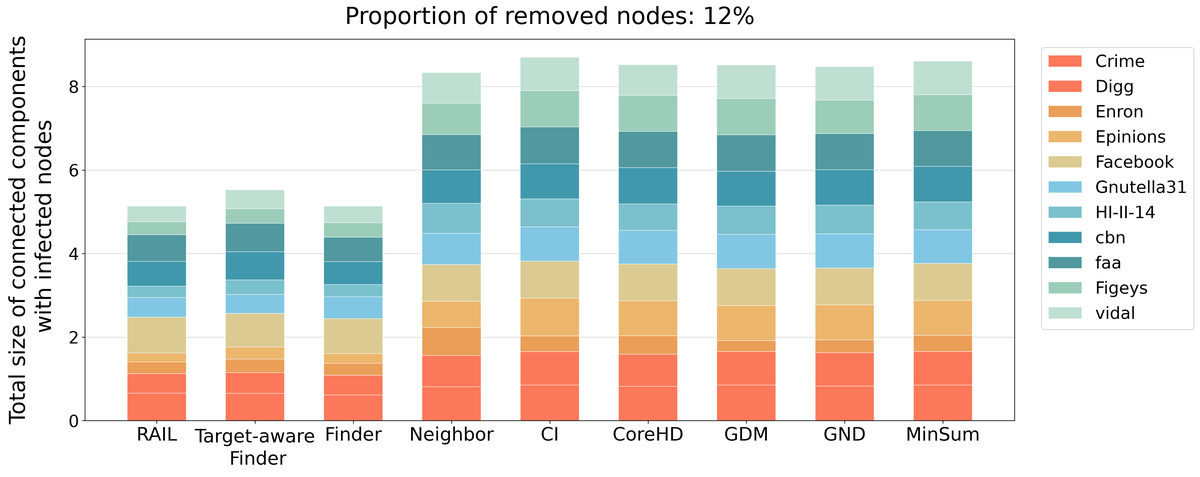}
        \caption{Removal ratio: 12\%}
        \label{fig:barfuzzy12}
    \end{subfigure}

    \vspace{0.4cm}

    \begin{subfigure}[t]{0.48\linewidth}
        \centering
        \includegraphics[width=\linewidth]{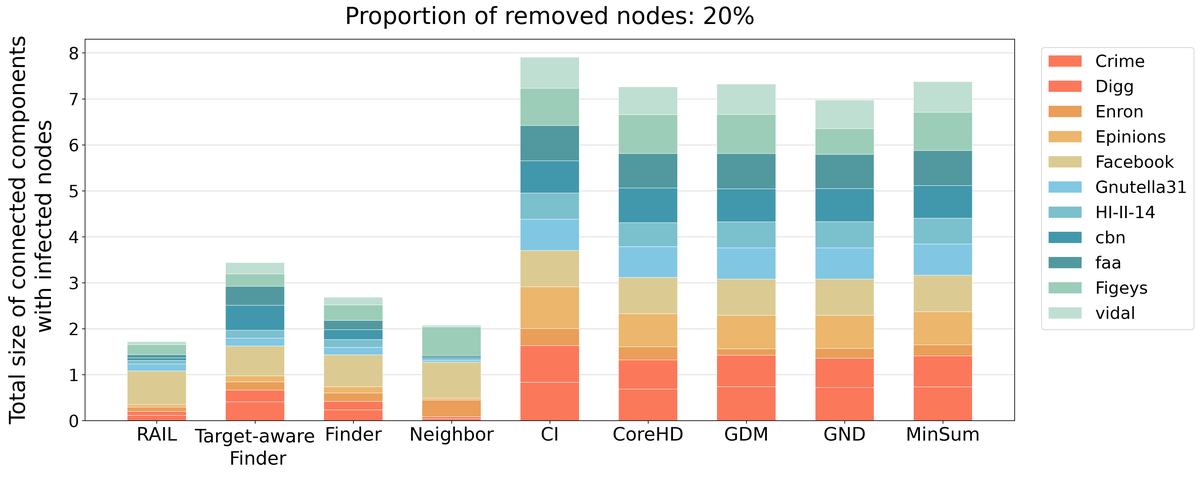}
        \caption{Removal ratio: 20\%}
        \label{fig:barfuzzy20}
    \end{subfigure}
    \hfill
    \begin{subfigure}[t]{0.48\linewidth}
        \centering
        \includegraphics[width=\linewidth]{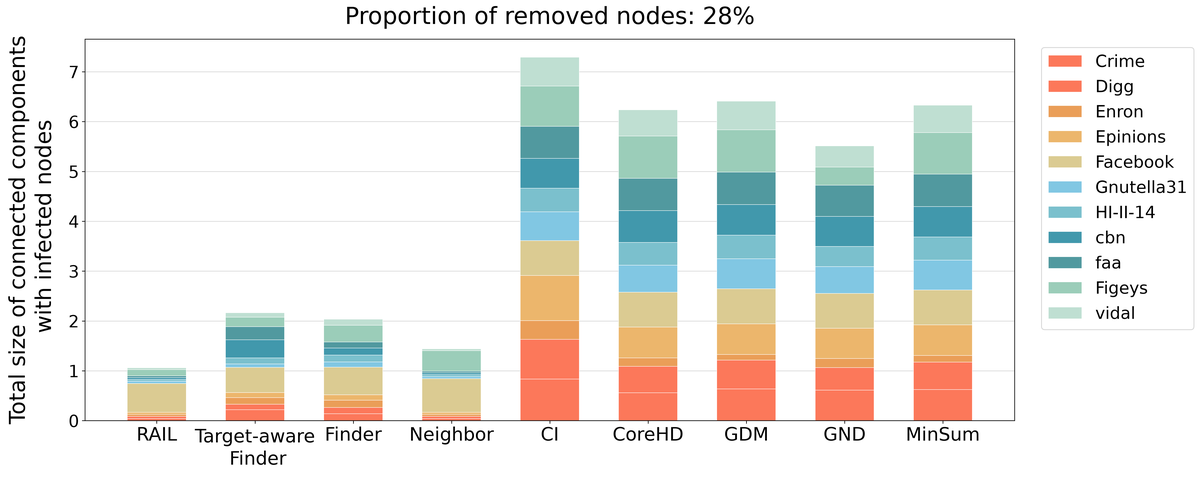}
        \caption{Removal ratio: 28\%}
        \label{fig:barfuzzy28}
    \end{subfigure}

    \caption{\textbf{Size of connected components containing at least one true infected node under uncertain infected nodes.} Each panel corresponds to one removal ratio. In this setting, each method observes an uncertain set of infected nodes, whereas the true infected nodes used for evaluation are sampled according to the assigned probabilities. For each method, the stacked bar aggregates the normalized $R_T(V_R)$ across all real networks. Smaller total bar height indicates better isolation performance.}
    \label{fig:barfuzzy}
\end{figure}

Figure~\ref{fig:bar} summarizes the results under fully observed infected nodes. At the low removal budget of $4\%$, RAIL performs comparably to the strongest network dismantling methods while substantially outperforming the Neighbor heuristic. This indicates that early intervention benefits from identifying structurally important nodes rather than simply removing neighbors of infected nodes. As the removal budget increases, RAIL consistently achieves the smallest residual $R_T(V_R)$, outperforming both network dismantling methods and the method Neighbor.  The results show that RAIL can adapt its node selections as the network evolves, consistently choosing nodes that more effectively isolate the infected nodes.

\begin{figure}[!htbp]
    \centering

    \begin{subfigure}[t]{0.43\linewidth}
        \centering
        \includegraphics[width=\linewidth]{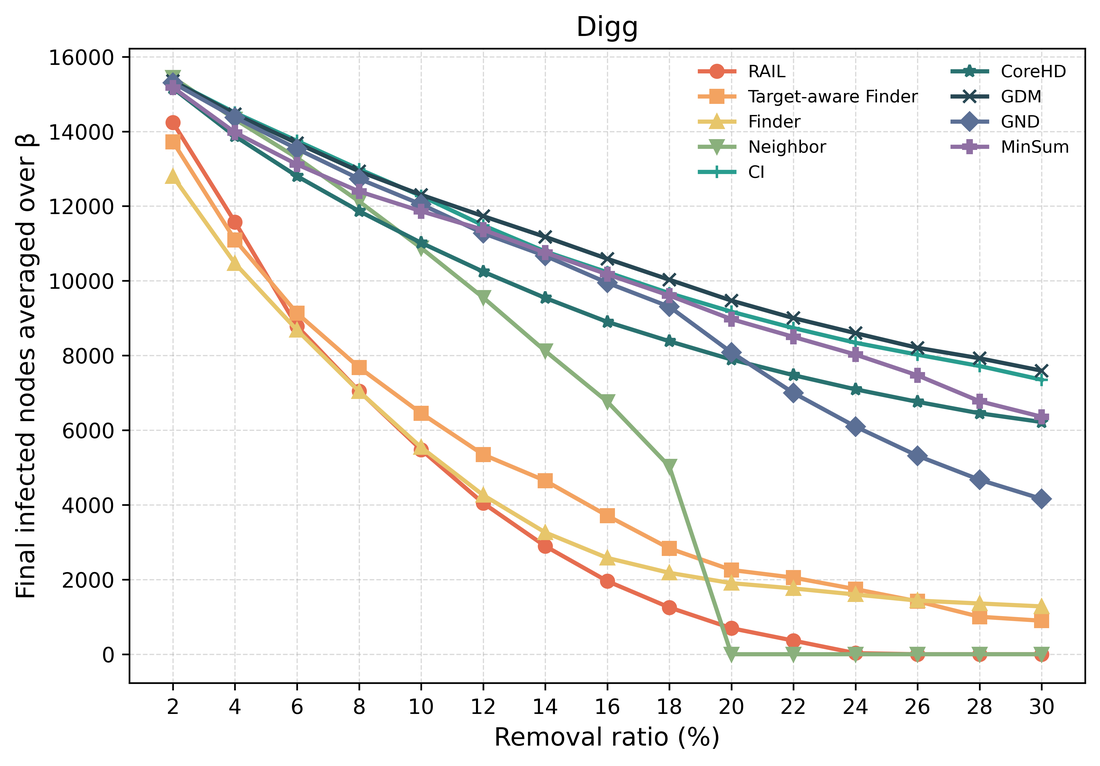}
        \caption{Digg}
        \label{fig:linedigg}
    \end{subfigure}
    \hfill
    \begin{subfigure}[t]{0.43\linewidth}
        \centering
        \includegraphics[width=\linewidth]{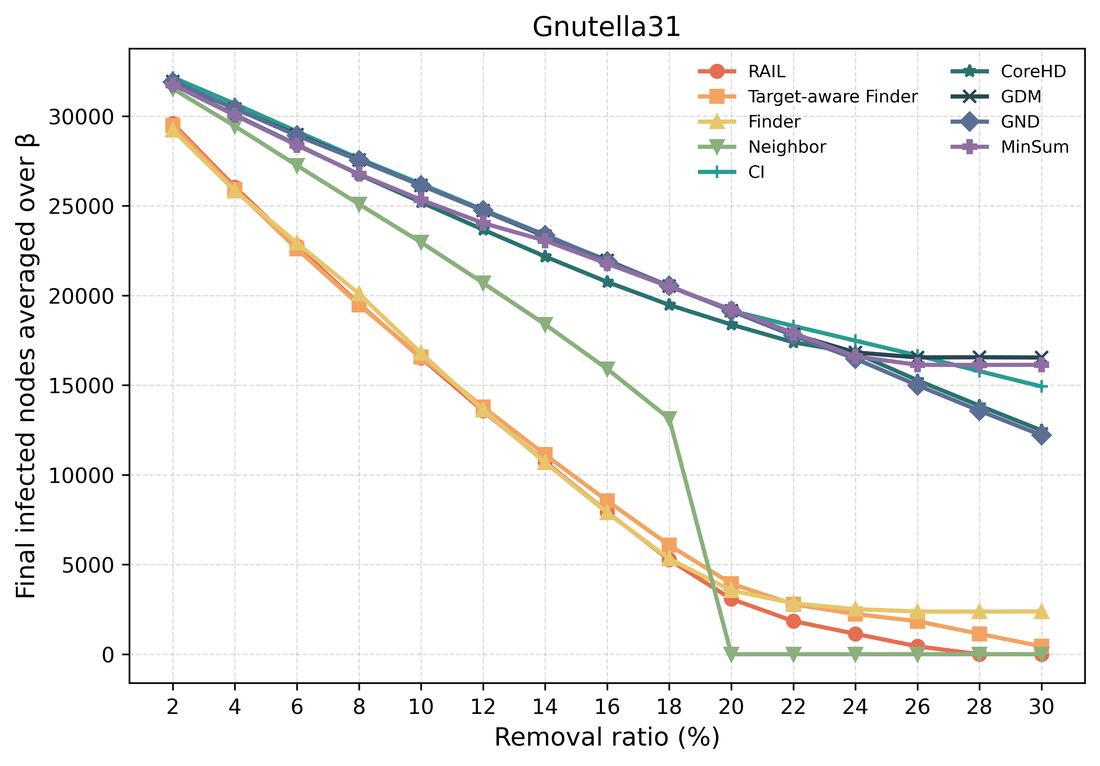}
        \caption{Gnutella31}
        \label{fig:lineG}
    \end{subfigure}

    \vspace{0.15cm}

    \begin{subfigure}[t]{0.43\linewidth}
        \centering
        \includegraphics[width=\linewidth]{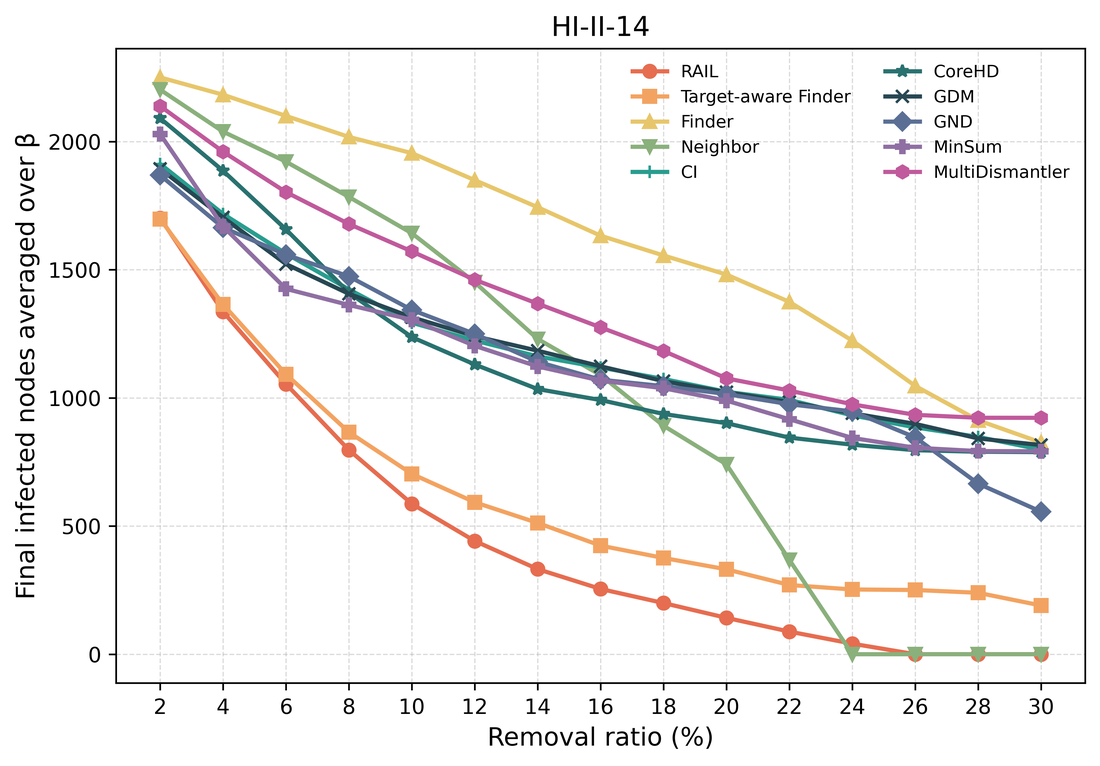}
        \caption{HI-II-14}
        \label{fig:lineH}
    \end{subfigure}
    \hfill
    \begin{subfigure}[t]{0.43\linewidth}
        \centering
        \includegraphics[width=\linewidth]{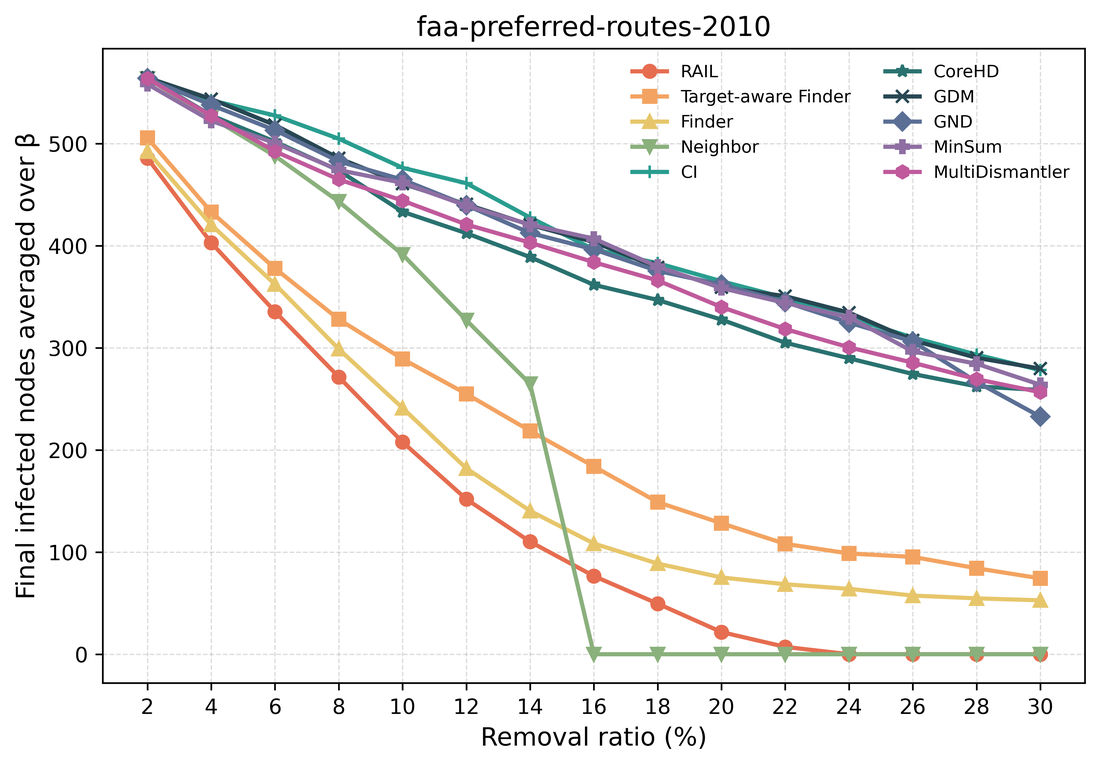}
        \caption{faa}
        \label{fig:linefaa}
    \end{subfigure}

    \vspace{0.15cm}

    \begin{subfigure}[t]{0.43\linewidth}
        \centering
        \includegraphics[width=\linewidth]{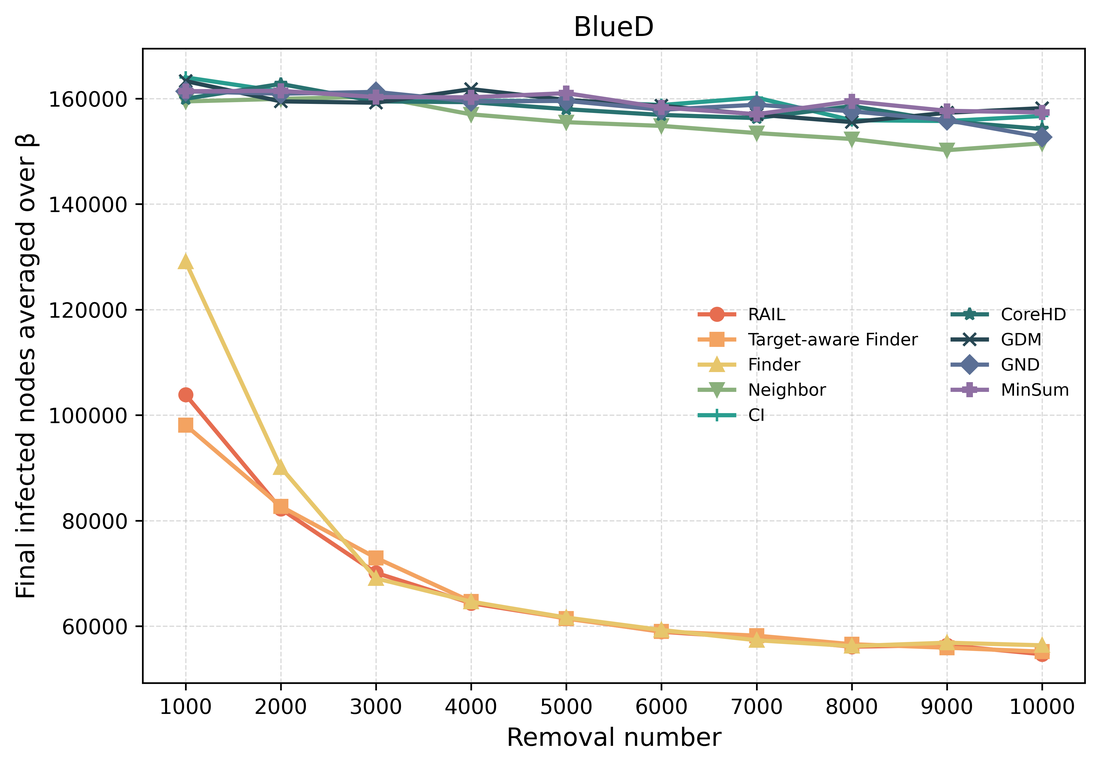}
        \caption{BlueD}
        \label{fig:lineblued}
    \end{subfigure}
    \hfill
    \begin{subfigure}[t]{0.43\linewidth}
        \centering
        \includegraphics[width=\linewidth]{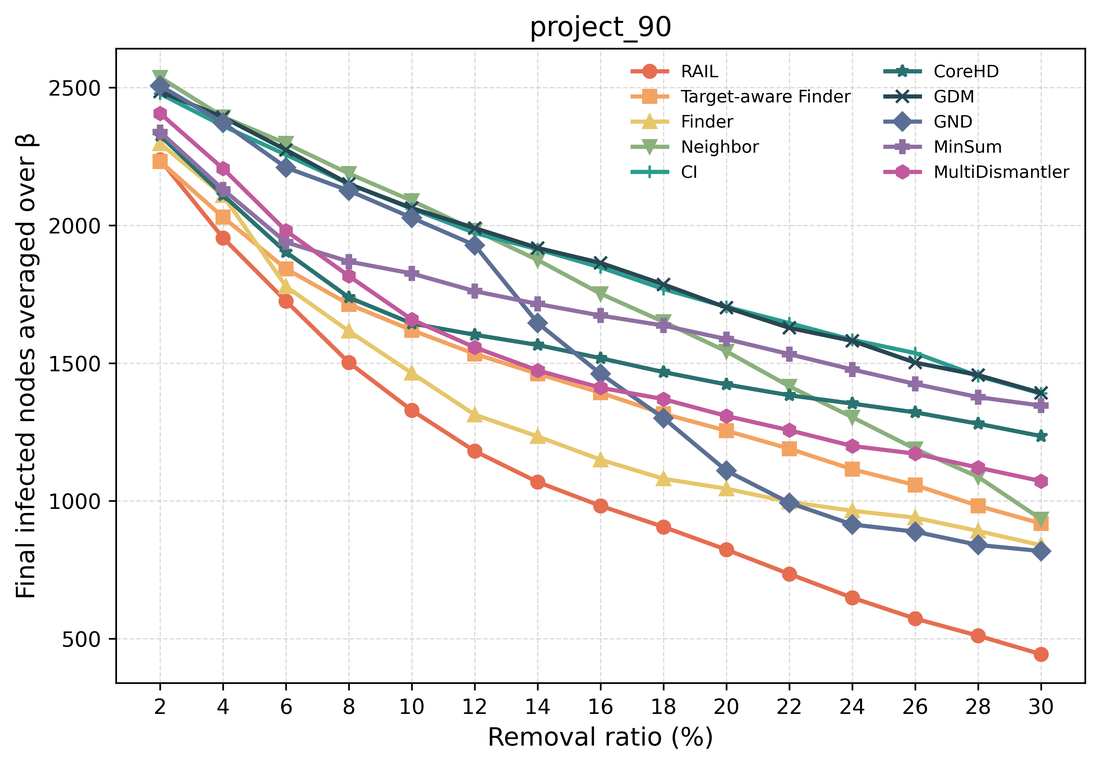}
        \caption{Project 90}
        \label{fig:lineproject}
    \end{subfigure}

    \caption{\textbf{Spreading suppression under SIR dynamics.}
    Panels (a)--(d) show representative real networks with randomly assigned infected nodes.
    Panels (e)--(f) show datasets in which infected nodes are predefined by the data.
    For each method, the curve shows the final number of infected nodes averaged over six
    infection probabilities after node removal. Smaller values indicate stronger spreading
    suppression.}

    \label{fig:sir_all}
\end{figure}

We further evaluate the same task under uncertain infected nodes. The algorithm observes a set of uncertain infected nodes, while evaluation is performed on nodes sampled from this set and treated as the infected nodes. As shown in Figure~\ref{fig:barfuzzy}, RAIL consistently outperforms the baselines across removal budgets, demonstrating robustness to imperfect identification of infected nodes.

\subsection{Suppressing Spreading from Infected Nodes}
We next test whether reducing the size of components containing infected nodes also suppresses spreading initiated from these nodes.

\subsubsection{Spreading Suppression in SIR Simulations}
We next test whether reducing these components also suppresses epidemic spreading from infected nodes. Figure~\ref{fig:sir_all}(a)-(d) shows representative results on four real networks, while the complete experimental settings and results for all datasets are provided in the Supplementary Section 6.

At small removal budgets, RAIL performs as well as the strongest network dismantling methods and substantially better than Neighbor. At larger budgets, its performance becomes comparable to Neighbor. Overall, RAIL achieves better and more robust performance than Neighbor across different removal budgets.





We further evaluate RAIL on two datasets, Blued and Project 90, in which the infected nodes are predefined by the data. As shown in Figure~\ref{fig:sir_all}(e)-(f), RAIL also achieves the strongest suppression when infected nodes are predefined rather than randomly assigned.



\begin{figure}[htbp]
    \centering

    \begin{subfigure}[t]{0.48\linewidth}
        \centering
        \includegraphics[width=\linewidth]{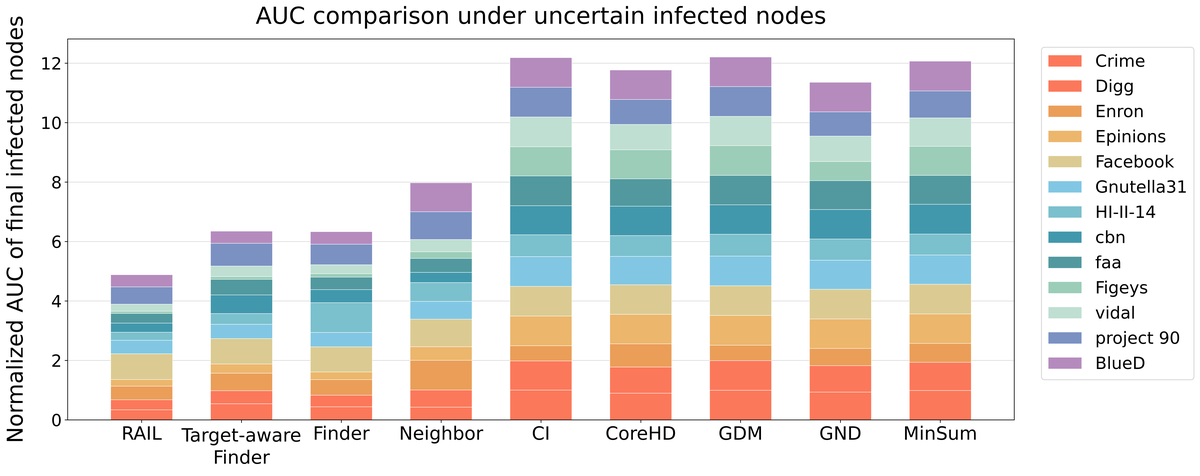}
        \caption{Fully observed infected nodes}
        \label{fig:baraucknown}
    \end{subfigure}
    \hfill
    \begin{subfigure}[t]{0.48\linewidth}
        \centering
        \includegraphics[width=\linewidth]{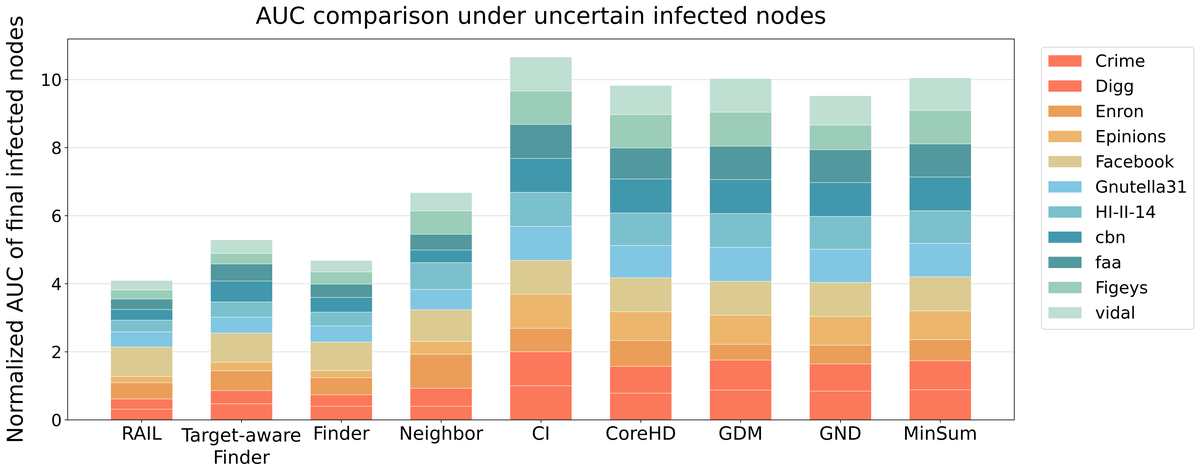}
        \caption{Uncertain infected nodes}
        \label{fig:baruncertain}
    \end{subfigure}

    \caption{\textbf{Aggregated normalized AUC for SIR suppression.} Results are shown under fully observed and uncertain infected nodes. For each dataset, the AUC obtained by each method is normalized. The normalized AUC values are then stacked across datasets for each method, with smaller total bar heights indicating stronger overall spreading suppression.}
    \label{fig:barauc}
\end{figure}

To provide an overall comparison across datasets, we summarize the SIR results using the normalized area under the removal–infection curve (AUC) for each method (see Supplementary Section 5 for details). As shown in Figure~\ref{fig:barauc}, RAIL achieves the smallest aggregated AUC under both fully observed and uncertain infected nodes, showing the strongest overall spreading suppression across datasets. Neighbor yields a larger aggregated AUC across datasets, indicating that removing only nodes adjacent to infected nodes is less effective overall.

\subsubsection{Spreading Suppression under Empirical Epidemic Arrival Sequences}

Beyond SIR simulations, we further evaluate RAIL using historical epidemic data on the global airport network~\cite{OAG}. We consider SARS~\cite{sarsh1n1}, H1N1~\cite{sarsh1n1}, and COVID-19~\cite{covid}, and construct a diffusion process for each outbreak according to the recorded order in which the disease reached different countries. Details are provided in Supplementary Section 5. 

As shown in Figure~\ref{fig:linereal}, RAIL achieves strong spreading suppression at small removal budgets and remains competitive as the removal budget increases. Although Neighbor becomes more effective after more nodes are removed, RAIL performs better across a wider range of removal budgets.

\begin{figure}[htbp]
    \centering

    \begin{subfigure}[t]{0.32\linewidth}
        \centering
        \includegraphics[width=\linewidth]{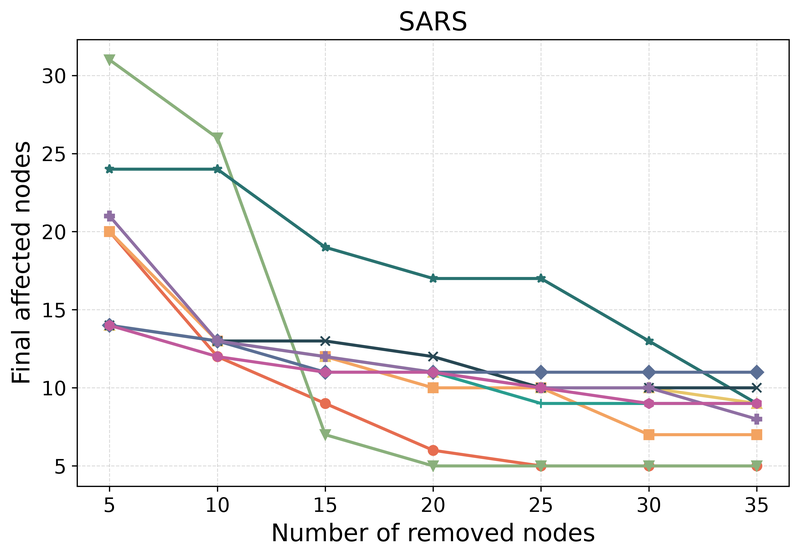}
        \caption{SARS}
        \label{fig:linesars}
    \end{subfigure}
    \hfill
    \begin{subfigure}[t]{0.32\linewidth}
        \centering
        \includegraphics[width=\linewidth]{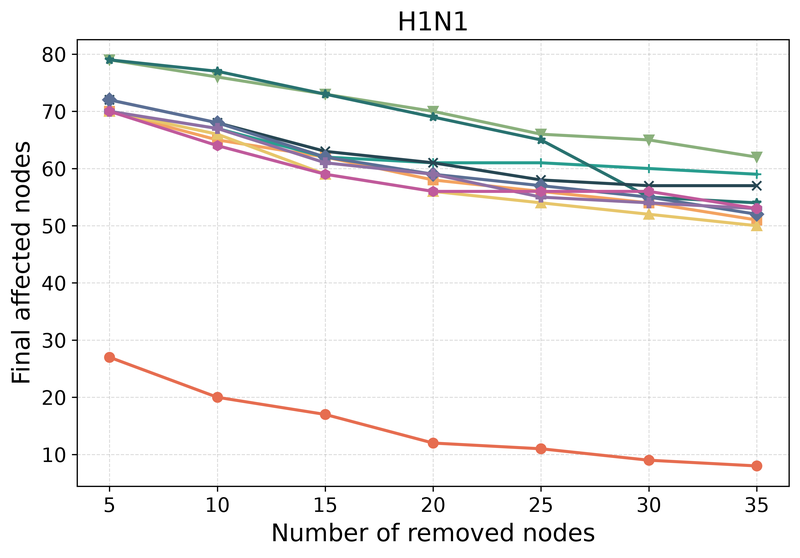}
        \caption{H1N1}
        \label{fig:lineh1n1}
    \end{subfigure}
    \hfill
    \begin{subfigure}[t]{0.32\linewidth}
        \centering
        \includegraphics[width=\linewidth]{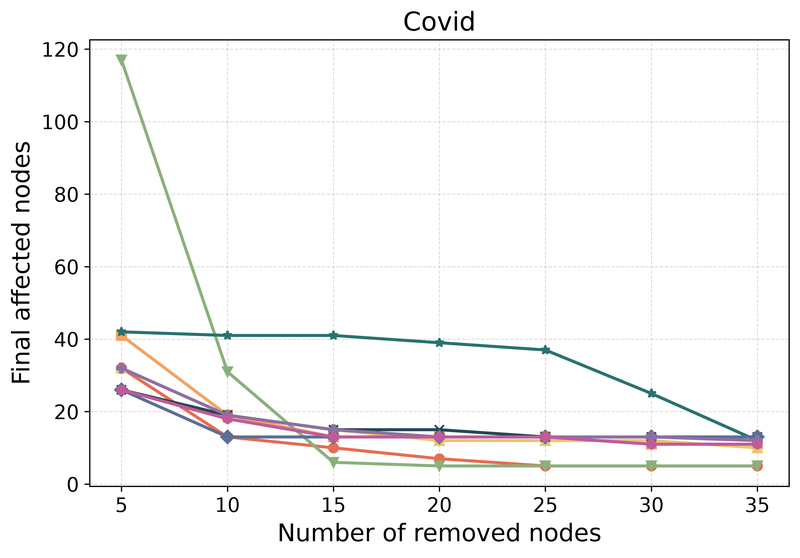}
        \caption{Covid-19}
        \label{fig:linecovid}
    \end{subfigure}

    \caption{\textbf{Spreading suppression under empirical epidemic arrival sequences.} Each panel shows one outbreak. The curves show the number of countries that remain connected through the airport network according to the observed order of epidemic arrival after node removal. Smaller values indicate stronger spreading suppression. The colors assigned to the methods are consistent with those in Figure~\ref{fig:sir_all}.}
    \label{fig:linereal}
\end{figure}

\section{Conclusions}

In this work, we fundamentally redefine network intervention under resource constraints, transitioning from the classical objective of global network fragmentation to the precise isolation of localized risks. By formulating target-aware network dismantling, we demonstrate that protecting a system from known threats requires dynamically adapting to the evolving residual network. Our analysis uncovers a critical budget-dependent transition, proving that the most effective intervention strategy is never static. To operationalize this theoretical insight, we developed RAIL, a deep reinforcement learning framework that continuously learns and adapts to the changing network topology. Extensive evaluations confirm that RAIL effectively minimizes the impact of infected components and robustly suppresses spreading dynamics, even under high uncertainty.

While our findings establish a robust foundation for dynamic risk isolation, they also unveil critical frontiers for future research. A compelling theoretical challenge lies in developing analytical or numerical methods to exactly determine the transition points ($R_c$) between structural-centric and neighbor-centric strategies across diverse network topologies. Furthermore, real-world interventions must navigate profound complexities: infected nodes may be underreported or dynamically shifting, and the underlying social, transportation, or biological networks continuously evolve over time. Expanding our framework to natively incorporate such spatiotemporal uncertainties, along with heterogeneous intervention costs and the necessity of preserving normal system functions, will bridge the gap between theoretical network science and complex operational deployments.

Crucially, our framework transcends theoretical algorithmic design to offer actionable, translational solutions for real-world public health crises. In the context of infectious disease control, the budget-dependent transition identified here provides a mathematically grounded rationale for adaptive policymaking. When diagnostic testing, quarantine facilities, or intervention personnel are severely limited, RAIL can be directly translated into targeted containment protocols. For instance, rather than exhausting scarce resources on traditional, exhaustive contact tracing, health authorities can leverage this framework to prioritize the isolation of structurally central individuals or key mobility hubs, thereby halting transmission cascades. By integrating our approach with real-time epidemiological data---such as global airport networks or local mobility tracking---policymakers can dynamically allocate resources and adjust travel restrictions as an outbreak evolves. Ultimately, this shifts epidemic management from static, reactive measures to an adaptive, forward-looking paradigm, maximizing population protection under strict operational constraints.

More broadly, this research catalyzes a paradigm shift in the study of network robustness and dismantling. By moving beyond blind global fragmentation to address localized, partially observed risks under strict resource limitations, we provide a generalizable framework for adaptive decision-making in evolving environments. By unifying complex network topology, observable risk metrics, and sequential deep reinforcement learning, our work not only advances the theoretical frontiers of network interventions but also opens new, highly pragmatic pathways for epidemic control, infrastructure protection, and the mitigation of cascading failures across interconnected systems.

\newpage
\noindent \textbf{Author contribution}\\
\noindent J.S. and Q.Z. conceptualized the study. Q.Z. and X.Q. supervised the study. J.S. developed the model and algorithm. J.S. and F.J. did the theoretical analysis. W.H. provided critical comments. J.S. and Q.Z. wrote the manuscript. All authors edited the manuscript.\\

\noindent \textbf{Data, ethics, and code statement}\\
\noindent This study utilized publicly available, de-identified, and aggregated data, as cited in the manuscript. Ethical approval was not required for this retrospective analysis. The code is available at \href{https://github.com/jkS529/RAIL}{https://github.com/jkS529/RAIL}.\\

\noindent\textbf{Acknowledgment}\\
\noindent The work was supported by the General Research Fund of the Research Grants Council of Hong Kong (No. 17209225, No. 17210226), the Shenzhen Medical Research Fund of the Shenzhen Medical Academy of Research and Translation (No. B032608), and Shenzhen Loop Area Institute (No. FPF10120250014 and No. SPF10120250004). \\

\noindent\textbf{Competing interest statement}\\
\noindent None.

\bibliographystyle{unsrt}
\bibliography{ref}
\end{document}